\documentclass[review]{elsarticle}
\usepackage{graphicx}
\usepackage{amsmath}
\usepackage{subfigure}
\usepackage{amsthm}
\usepackage{amssymb}
\usepackage[margin=1.65cm,nohead]{geometry}

\begin{document}

\begin{frontmatter}

\title{Stability and Performance Issues of a Relay Assisted Multiple Access Scheme with MPR Capabilities}

\tnotetext[t1]{This work was presented in part at the 9th Intl. Symposium on Modeling and Optimization in Mobile, Ad Hoc, and Wireless Networks, Princeton USA, May 2011~\cite{b:Pappas-WiOpt}. This work was performed when N.Pappas was with Computer Science Department, University of Crete, Greece and the Institute of Computer Science, Foundation for Research and Technology - Hellas (FORTH). }

\author{Nikolaos Pappas\fnref{fn1}}
\ead{nikolaos.pappas@supelec.fr}
\address{Sup\'{e}lec, Department of Telecommunications, Gif-sur-Yvette, France}

\author{Anthony Ephremides}
\ead{etony@umd.edu}
\address{Department of Electrical and Computer Engineering and Institute for Systems Research\\ University of Maryland, College Park, MD 20742}

\author{Apostolos Traganitis}
\ead{tragani@ics.forth.gr}
\address{Computer Science Department, University of Crete, Greece\\Institute of Computer Science, Foundation for Research and Technology - Hellas (FORTH)}

\fntext[fn1]{Corresponding author}

\maketitle

\begin{abstract}
In this work, we study the impact of a relay node to a network with a finite number of users-sources and a destination node. We assume that the users have saturated queues and the relay node does not have packets of its own; we have random access of the medium and the time is slotted. The relay node stores a source packet that it receives successfully in its queue when the transmission to the destination node has failed. The relay and the destination nodes have multi-packet reception capabilities. We obtain analytical equations for the characteristics of the relay's queue such as average queue length, stability conditions etc. We also study the throughput per user and the aggregate throughput for the network.
\end{abstract}

\begin{keyword}
Relay \sep MPR \sep Multiple Access \sep Stability
\end{keyword}

\end{frontmatter}

\section{Introduction}
\label{sec:intro}

In this work we examine the operation of a node relaying packets from a number of users-sources to a destination node as shown in Fig.~\ref{fig:netmodel}, and is an extension of our work in~\cite{b:Pappas} (in that work we assumed random access scheme with collision channel model with erasures). We assume multi-packet reception (MPR) capability for the relay and the destination node.

The classical relay channel was originally introduced by van der Meulen~\cite{b:Muelen}. Earlier works on the relay channel were based on information theoretical formulations as in~\cite{b:CoverGamal} and~\cite{b:Sadek}. Recently several works have investigated relaying capability at the MAC layer~\cite{b:Sadek, b:Simeone, b:Rong1, b:Rong2, b:Pappas-ITW11-Relay, b:Pappas-ISIT12, b:Pappas-arXiv-full-duplex}. More specifically, in~\cite{b:Sadek}, the authors have studied the impact of cooperative communications at the multiple-access layer with TDMA. They introduced a new cognitive multiple-access protocol in the presence of a relay in the network. The relay senses the channel for idle channel resources and exploits them to cooperate with the terminals in forwarding their packets. Most cooperative techniques studied so far have been on physical layer cooperation, however there is evidence (as in~\cite{b:Sadek}) that the same gains can be achieved with network layer cooperation, that is plain relaying without any physical layer considerations.

The classical analysis of random multiple access schemes like slotted ALOHA~\cite{b:Bertsekas} has focused on the collision model, the collision channel however is not the appropriate for wireless networks. Random access with MPR has attracted attention recently~\cite{b:AlohaVerdu},~\cite{b:Angel}. The authors in~\cite{b:Naware} consider the effect of MPR on stability and delay of slotted ALOHA based random-access system and it is shown that the stability region undergoes a phase transition from a concave region to a convex polyhedral region as the MPR capability improves. All these previous approaches come together in the model that we consider in this paper.

We assume random access to the channel, time is considered slotted, and each packet transmission takes one time slot. The wireless channel between the nodes in the network is modeled by a Rayleigh narrowband flat-fading channel with additive Gaussian noise. The relay and the destination are equipped with multiuser detectors, so that they may decode packets successfully from more than one transmitter at a time (MPR capability). A user's transmission is successful if the received signal to interference plus noise ratio ($SINR$) is above a threshold $\gamma$. We also assume that acknowledgements (ACKs) are instantaneous and error free. The relay does not have packets of its own and the sources are considered saturated with unlimited amount of traffic.

We obtain analytical expressions for the characteristics of the relay's queue such as arrival and service rate of the relay's queue, the stability condition and the average length of the queue as functions of the probabilities of transmissions and the outage probabilities of the links. We study the impact of the relay node on the throughput per user-source and the aggregate throughput. We show that the throughput per user-source does not depend on the probability of the relay transmission and that there is an optimum number of users that maximizes the aggregate throughput.

Section~\ref{sec:sysmod} describes the system model, in Section~\ref{sec:analysis} we study the characteristics of the relay's queue and we derive the equations for the throughput per user and the aggregate throughput. We present the arithmetic and simulation results in Section~\ref{sec:results} and, finally, our conclusions are given in Section~\ref{sec:conclusions}.

\section{System Model}
\label{sec:sysmod}
\subsection{Network Model}
\begin{center}
\begin{figure}[ht]
\begin{minipage}[b]{0.3\linewidth}
\centering
\includegraphics[scale=0.45]{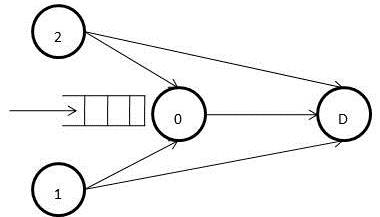}
\caption{The simple network model}
\label{fig:netmodel}
\end{minipage}
\begin{minipage}[b]{0.3\linewidth}
\centering
\includegraphics[scale=0.4]{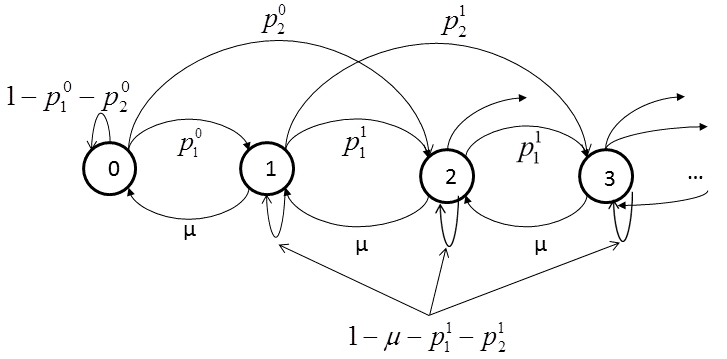}
\caption{Markov Chain model}
\label{fig:mc}
\end{minipage}
\end{figure}
\end{center}

We consider a network with $N$ users-sources, one relay node and a single destination node. The sources transmit packets to the destination with the cooperation of the relay; the case of $N=2$ is depicted in Fig.~\ref{fig:netmodel}. We assume that the queues of the two sources are saturated (i.e. there are no external arrivals); the relay does not have packets of its own, and just forwards the packets that it has received from the two users. The relay node stores a source packet that it receives successfully in its queue when the direct transmission to the destination node has failed. We assume random access of the medium. Each of the receivers (relay and destination) is equipped with multiuser detectors, so that they may decode packets successfully from more than one transmitter at a time. Nodes cannot transmit and receive at the same time. The queue in the relay has infinite capacity.

It is important to note that the relay node must be easier accessible than the destination, meaning that the user - relay channel has to be more reliable than
the user-destination one. At the same time the relay - destination channel must be more reliable than the user - destination channel. Otherwise the presence of the relay degrades the performance of the whole network. In the following subsection we present all the details about the physical layer model assumed in this work.

\subsection{Physical Layer Model}
The MPR channel model used in this paper is a generalized form of the packet erasure model. In the wireless environment, a packet can be decoded correctly by the receiver if the received $SINR$ exceeds a certain threshold. More precisely, suppose that we are given a set $T$ of nodes transmitting in the same time slot. Let  $P_{rx}(i,j)$ be the signal power received from node $i$ at node $j$ (when $i$ transmits), and let $SINR(i,j)$ be the $SINR$ determined by node $j$, i.e.,

\begin{equation*}
SINR(i,j)=\frac{P_{rx}(i,j)}{\eta_{j}+\sum_{k\in T\backslash\left\{i\right\}} {P_{rx}(k,j)}}
\end{equation*}

where $\eta_{j}$ denotes the receiver noise power at $j$. We assume that a packet transmitted by $i$ is successfully received by $j$ if and only if $SINR(i,j)\geq \gamma_{j}$, where $\gamma_{j}$ is a threshold characteristic of node $j$. The wireless channel is subject to fading; let $P_{tx}(i)$ be the transmitting power at node $i$ and $r(i,j)$ be the distance between $i$ and $j$. The power received by $j$ when $i$ transmits is $P_{rx}(i,j)=A(i,j)g(i,j)$ where $A(i,j)$ is a random variable representing channel fading. We assume that the fading model is slow, flat fading, constant during a timeslot and independently varying from timeslot to timeslot. Under Rayleigh fading, it is known~\cite{b:Tse} that $A(i,j)$ is exponentially distributed. The received power factor $g(i,j)$ is given by $g(i,j)=P_{tx}(i)(r(i,j))^{-\alpha}$ where $\alpha$ is the path loss exponent with typical values between $2$ and $4$. The success probability of link $ij$ when the transmitting nodes are in $T$ is given by

\begin{equation}
\label{eq:succprob}
P_{i/T}^{j}=\exp\left(-\frac{\gamma_{j}\eta_{j}}{v(i,j)g(i,j)}\right) \prod_{k\in T\backslash \left\{i,j\right\}}{\left(1+\gamma_{j}\frac{v(k,j)g(k,j)}{v(i,j)g(i,j)}\right)}^{-1}
\end{equation}

where $v(i,j)$ is the parameter of the Rayleigh random variable for fading. The analytical derivation for this success probability can be found in~\cite{b:Nguyen}.

\section{Analysis}
\label{sec:analysis}
In this section we derive the equations for the characteristics of the relay's queue, such as the arrival and service rates, the stability conditions, and the average queue length. We provide an analysis for two cases: first, when the network consists of two users (non-symmetric) and the second is for $n>2$ symmetric users.
\subsection{Two-user case}
The service rate is given by
\begin{equation}
\begin{aligned}
\mu=q_{0}(1-q_{1})(1-q_{2})P_{0/0}^{d}+q_{0}q_{1}(1-q_{2})P_{0/0,1}^{d}
+q_{0}q_{2}(1-q_{1})P_{0/0,2}^{d}+q_{0}q_{1}q_{2}P_{0/0,1,2}^{d}
\end{aligned}
\end{equation}

where $q_{0}$ is the transmission probability of the relay given that it has packets in its queue, $q_{i}$ for $i\neq0$ is the transmission probability for the $i$-th user.The term $P_{i/i,k}^{j}$ is the success probability of link $ij$ when the transmitting nodes are $i$ and $k$ and is given by (\ref{eq:succprob}).

If the queue of the relay node is empty, the arrival rate is denoted by $\lambda_{0}$ and if it is not by $\lambda_{1}$.

In order to study the stability of the queue, we will present the queue dynamics. The dynamics of the queue at a timeslot $t$ are given by:

\begin{equation}
Q(t+1) = \left[ Q(t) - Y(t)\right]^+ + X(t)
\end{equation}

Where $[x]^+ = \max(x,0)$, $Q(t)$ is the queue size at $t$, $Y(t)$ denotes the departure process of a packet with $\mathbb{E} \lbrace Y(t)\rbrace=\mu$. Furthermore $X(t)$ denotes the arrival process of the packets, if $Q(t)=0$ then $\mathbb{E} \lbrace X(t)\rbrace=\lambda_0$, otherwise $\mathbb{E} \lbrace X(t)\rbrace=\lambda_1$.

Define the Lyapunov function~\cite{b:NOW} $L(Q(t))$ as follows:

\begin{equation}
L(Q(t))=Q(t)
\end{equation}

Given that $Q(t)>1$ at a timeslot $t$ the Lyapunov Drift~\cite{b:NOW} is:

\begin{equation}
\Delta (Q(t)) \triangleq \mathbb{E} \lbrace L(Q(t+1))-L(Q(t)) \vert Q(t) \rbrace
\end{equation}

\begin{align*}
\Delta (Q(t)) =  \mathbb{E} \lbrace \left[ Q(t) - Y(t)\right]^+ \vert Q(t) \rbrace + \mathbb{E} \lbrace X(t) \vert Q(t) \rbrace - Q(t) = \\
= \mathbb{E} \lbrace \left[ Q(t) - Y(t)\right]^+ \vert Q(t) \rbrace + \lambda_1 - Q(t)= \\
= \mathrm{Pr} [Q(t) > Y(t)] \mathbb{E} \lbrace Q(t) - Y(t) \vert Q(t) , Q(t) > Y(t) \rbrace + \lambda_1 - Q(t) \\
\Delta (Q(t)) \leq \mathbb{E} \lbrace Q(t) - Y(t) \vert Q(t) , Q(t) > Y(t) \rbrace + \lambda_1 - Q(t)= \\
=\mathbb{E} \lbrace Q(t) \vert Q(t) , Q(t) > Y(t) \rbrace +\mathbb{E} \lbrace - Y(t) \vert Q(t) , Q(t) > Y(t)\rbrace+ \lambda_1 - Q(t)=\\
=\mathbb{E} \lbrace Q(t) \vert Q(t) , Q(t) > Y(t) \rbrace - \mu + \lambda_1 - Q(t)
\end{align*}

Thus we have that $\Delta (Q(t)) \leq - \mu + \lambda_1$. The queue is strongly stable~\cite{b:NOW} if $\lambda_1 < \mu$.

Given that the queue is stable if $\lambda_1 < \mu$, we have the following equation for the average arrival rate $\lambda$
\begin{equation}
\lambda=P\left(Q=0\right)\lambda_{0}+P\left(Q>0\right)\lambda_{1}
\end{equation}

If the queue of the relay is empty then the relay, naturally, does not attempt to transmit, thus the probability of arrival is $\lambda_{0}=p_{1}^{0}+2p_{2}^{0}$, where $p_{i}^{0}$ is the probability of receiving $i$ packets given that the queue is empty. The expressions for $p_{i}^{0}$ are:
\begin{equation}
\begin{aligned}
p_{1}^{0}=q_{1}(1-q_{2})(1-P_{1/1}^{d})P_{1/1}^{0}+q_{2}(1-q_{1})(1-P_{2/2}^{d})P_{2/2}^{0}
+q_{1}q_{2}(1-P_{1/1,2}^{d})P_{1/1,2}^{0}\left[P_{2/1,2}^{d}+(1-P_{2/1,2}^{d})(1-P_{2/1,2}^{0})\right]+\\
+q_{1}q_{2}(1-P_{2/1,2}^{d})P_{2/1,2}^{0}\left[P_{1/1,2}^{d}+(1-P_{1/1,2}^{d})(1-P_{1/1,2}^{0})\right]
\end{aligned}
\end{equation}

\begin{equation}
p_{2}^{0}=q_{1}q_{2}(1-P_{1/1,2}^{d})(1-P_{2/1,2}^{d})P_{1/1,2}^{0}P_{2/1,2}^{0}
\end{equation}

If the queue is not empty then the arrival rate is given by $\lambda_{1}=p_{1}^{1}+2p_{2}^{1}$, where $p_{i}^{1}$ is the probability of receiving $i$ packets given that the queue is not empty and $p_{i}^{1}=(1-q_{0})p_{i}^{0}$; thus $\lambda_{1}=(1-q_{0})\lambda_{0}$, this is because the relay cannot receive and transmit at the same time. In Fig.~\ref{fig:mc}, we present the discrete time Markov Chain that describes the queue evolution. Each state, denoted by an integer, represents the queue size at the relay node. The transition matrix of the above DTMC is a lower Hessenberg matrix and is given by:

\begin{equation}
P=\left(\begin{array}{ccccc}
a_{0} & b_{0} & 0  & 0    & \cdots   \\
a_{1} & b_{1} & b_{0} & 0 & \cdots    \\
a_{2} & b_{2} & b_{1} & b_{0} & \cdots \\
0 & b_{3} & b_{2} & b_{1} & \cdots \\
0 & 0 & b_{3} & b_{2} & \cdots\\
\vdots & \vdots & \vdots & \vdots & \ddots
\end{array} \right)
\end{equation}

Where $a_{0}=1-p_{1}^{0}-p_{2}^{0}, a_{1}=p_{1}^{0}, a_{2}=p_{2}^{0}$ and
$b_{0}=\mu, b_{1}=1-\mu-p_{1}^{1}-p_{2}^{1}, b_{2}=p_{1}^{1}, b_{3}=p_{2}^{1}$.

Since the $P$ is an infinite-dimension matrix, we are going to obtain the expression for the steady-state distribution
vector $s$ using difference equations. The difference equations are given by:
\begin{equation}
Ps=s \Rightarrow s_{i}=a_{i}s_{0}+\sum_{j=1}^{i+1}{b_{i-j+1}s_{j}}
\end{equation}
We apply Z-transform technique to compute the steady-state distribution:
\begin{equation}
A(z)=\sum_{i=0}^{2}{a_{i}z^{-i}}, B(z)=\sum_{i=0}^{3}{b_{i}z^{-i}}, S(z)=\sum_{i=0}^{\infty}{s_{i}z^{-i}}
\end{equation}
It is known that~\cite{b:Gebali}:
\begin{equation}
S(z)=s_{0}\frac{z^{-1}A(z)-B(z)}{z^{-1}-B(z)}
\end{equation}
The probability that the queue in the relay is empty is given by the following formula~\cite{b:Gebali}:
\begin{equation}
\label{eq:s_0}
P\left(Q=0\right)=s_{0}=\frac{1+B^{'}(1)}{1+B^{'}(1)-A^{'}(1)}
\end{equation}
Where $A^{'}(1)=-p_{1}^{0}-2p_{2}^{0}$ and $B^{'}(1)=\mu-p_{1}^{1}-2p_{2}^{1}-1$. Then the probability that the queue in the relay is empty is given by:
\begin{equation}
P\left(Q=0\right)=\frac{\mu-\lambda_{1}}{\mu-\lambda_{1}+\lambda_{0}}
\end{equation}
From the above equations we can compute the average arrival rate $\lambda$:
\begin{equation}
\lambda=P\left(Q=0\right)\lambda_{0}+P\left(Q>0\right)\lambda_{1}=\frac{\mu\lambda_{0}}{\mu-\lambda_{1}+\lambda_{0}}
\end{equation}
Note that the average arrival rate does not depend on $q_{0}$ (the proof is straightforward and thus is omitted).

The queue is stable if $q_{0}$ satisfies $q_{0min}<q_{0}<1$. The expression for $q_{0min}$ is given by (\ref{eq:q0min2}), in order to
obtain $q_{0min}$ we have to solve the inequality $\lambda_1 < \mu$.

\begin{equation}
\label{eq:q0min2}
q_{0min}=\frac{\lambda_{0}}{\lambda_{0}+(1-q_{1})(1-q_{2})P_{0/0}^{d}+q_{1}(1-q_{2})P_{0/0,1}^{d}+q_{2}(1-q_{1})P_{0/0,2}^{d}+q_{1}q_{2}P_{0/0,1,2}^{d}}
\end{equation}

Notice that the conditions $\frac{\lambda}{\mu}<1$ and $\frac{\lambda_{1}}{\mu}<1$ are equivalent in our model.
\begin{equation*}
\begin{aligned}
\frac{\lambda}{\mu}<1 \Leftrightarrow \lambda<\mu \Leftrightarrow \frac{\mu\lambda_{0}}{\mu-\lambda_{1}+\lambda_{0}}<\mu \Leftrightarrow \frac{\lambda_{0}}{\mu-\lambda_{1}+\lambda_{0}}<1
\Leftrightarrow \lambda_{0}<\mu-\lambda_{1}+\lambda_{0} \Leftrightarrow \frac{\lambda_{1}}{\mu}<1
\end{aligned}
\end{equation*}

It is known~\cite{b:Gebali} that the average queue size is $\overline{Q}=-S^{'}(1)$, where $S^{'}(1)=s_{0}\frac{K^{''}(1)}{L^{''}(1)}$.
The expression for $K(z)$ is given by

\begin{equation}
\label{eq:Kz2}
K(z)=\left(-z^{-2}A(z)+z^{-1}A^{'}(z)-B^{'}(z) \right) \left(z^{-1}-B(z)\right) - \left(z^{-1}A(z)-B(z) \right) \left(-z^{-2}-B^{'}(z) \right)\\
\end{equation}

and $L(z)=\left(z^{-1}-B(z) \right)^{2}$.

After some algebra the average queue size is given by:
\begin{equation}
\overline{Q}=\frac{(\lambda_{1}-\mu)(2p_{1}^{0}+5p_{2}^{0})+\lambda_{0}(\mu-2p_{1}^{0}-5p_{2}^{0})}{(\mu-\lambda_{1}+\lambda_{0})(\lambda_{1}-\mu)}
\end{equation}

The throughput rate $\mu_{i}$ for the user $i$ is given by the (\ref{eq:mu1}) and (\ref{eq:mu2}).

\begin{equation}
\label{eq:mu1}
\begin{aligned}
\mu_{1}=q_{0}P\left(Q>0\right)q_{1}\left((1-q_{2})P_{1/0,1}^{d}+q_{2}P_{1/0,1,2}^{d} \right)+\\
+\left[1-q_{0}P\left(Q>0\right)\right]q_{1} \left[(1-q_{2})\left(P_{1/1}^{d}+(1-P_{1/1}^{d})P_{1/1}^{0} \right)+q_{2}\left(P_{1/1,2}^{d}+(1-P_{1/1,2}^{d})P_{1/1,2}^{0} \right) \right]
\end{aligned}
\end{equation}

\begin{equation}
\label{eq:mu2}
\begin{aligned}
\mu_{2}=q_{0}P\left(Q>0\right)q_{2}\left((1-q_{1})P_{2/0,2}^{d}+q_{1}P_{2/0,1,2}^{d} \right)+\\
+\left[1-q_{0}P\left(Q>0\right)\right]q_{2} \left[(1-q_{1})\left(P_{2/2}^{d}+(1-P_{2/2}^{d})P_{2/2}^{0} \right)+q_{1}\left(P_{2/1,2}^{d}+(1-P_{2/1,2}^{d})P_{2/1,2}^{0} \right) \right]
\end{aligned}
\end{equation}

In (\ref{eq:mu1}) and (\ref{eq:mu2}) we assume that the queue is stable, hence the arrival rate from each user to the queue is the contributed throughput from it. The aggregate throughput is $\mu_{total}=\mu_{1}+\mu_{2}$. Notice that the throughput per user is independent of $q_{0}$ as long as it is in the stability region. This is explained because the product $q_{0}P\left(Q>0\right)$ is constant. If we consider the previous network without the relay node then the throughput rates for the users are the following:

\begin{equation*}
\begin{aligned}
\mu_{1}=q_{1}(1-q_{2})P_{1/1}^{d}+q_{1}q_{2}P_{1/1,2}^{d}\\
\mu_{2}=q_{2}(1-q_{1})P_{2/2}^{d}+q_{1}q_{2}P_{2/1,2}^{d}
\end{aligned}
\end{equation*}

\subsection{N-symmetric users}
We now generalize the above for the case of a symmetric $n$-users network. Each user attempts to transmit in a slot with probability $q$; the success probability to the relay and the destination when $i$ nodes transmit are given by $P_{0,i}$, $P_{d,i}$  respectively. There are two cases for the $P_{d,i}$, $P_{d,i,0}$, $P_{d,i,1}$ denoting success probability when relay remains silent or transmits respectively. Finally $P_{0d,i}$ is the link probability of success from the relay to the destination when $i$ nodes transmit. The above success probabilities for the symmetric case are given by $P_{0,i}=P_{0}\left(\frac{1}{1+\gamma_{0}} \right)^{i-1}$, $P_{d,i,j}=P_{d}\left(\frac{1}{1+\gamma_{d}} \right)^{i-1} \left(\frac{1}{1+\beta\gamma_{0}} \right)^{j}$, $j=0,1$ and $\beta=\frac{v_{0d}g_{0d}}{v_{d}g_{d}}>1$. $P_{0d,i}=P_{0d}\left(\frac{1}{1+\frac{1}{\beta}\gamma_{d}}\right)^{i}$, $P_{0}=\exp\left(-\frac{\gamma_{0}\eta_{0}}{v_{0}g_{0}}\right)$, $P_{d}=\exp\left(-\frac{\gamma_{d}\eta_{d}}{v_{d}g_{d}}\right)$, $P_{0d}=\exp\left(-\frac{\gamma_{0}\eta_{0}}{v_{0}g_{0}}\right)$.

The service rate is given by the following equation:
\begin{equation}
\mu=\sum_{k=0}^{n}{{n \choose k} {q_{0}q^{k}(1-q)^{n-k}}P_{0d,k}}
\end{equation}
The average arrival rate $\lambda$ of the queue is given by:
\begin{equation}
\label{eq:lamdan}
\lambda=P\left(Q=0\right)\lambda_{0}+P\left(Q>0\right)\lambda_{1}
\end{equation}
Where $\lambda_{0}=\sum_{i=0}^{n}{i p_{i}^{0}}$ and $\lambda_{1}=(1-q_{0})\lambda_{0}$. $p_{i}^{0}$ is the probability of receiving $i$ packets given that the queue is empty, the expression for $p_{i}^{0}$ is given by (~\ref{eq:pk0n}). $p_{i}^{1}$ is the probability of receiving $i$ packets given that the queue is not empty and $p_{k}^{1}=(1-q_{0})p_{k}^{0}$.

\begin{equation}
\label{eq:pk0n}
p_{k}^{0}=\sum_{i=k}^{n}{{n \choose i}{i \choose k} {q^{i}(1-q)^{n-i}}P_{0,i}^{k}\left(1-P_{d,i,0}\right)^{k}\left[1-P_{0,i}(1-P_{d,i,0})\right]^{i-k}},\text{ }1 \leq k \leq n
\end{equation}

The elements of the transition matrix are $a_{0}=1-\sum_{i=1}^{n}{p_{i}^{0}}$, $a_{i}=p_{i}^{0}$  $\forall i>0$ and
$b_{0}=\mu$, $b_{1}=1-\mu-\sum_{i=1}^{n}{p_{i}^{1}}$,$b_{i+1}=p_{i}^{1}$  $\forall i>1$. The Z-transforms are:

\begin{equation}
A(z)=\sum_{i=0}^{n}{a_{i}z^{-i}}, B(z)=\sum_{i=0}^{n+1}{b_{i}z^{-i}}, S(z)=\sum_{i=0}^{\infty}{s_{i}z^{-i}}
\end{equation}

Following the same methodology as in the two-user case and applying the above to (\ref{eq:s_0}) we obtain the probability that the queue in the relay is empty is given by:
\begin{equation}
P\left(Q=0\right)=\frac{\mu-\lambda_{1}}{\mu-\lambda_{1}+\lambda_{0}}
\end{equation}

The queue is stable if $q_{0}$ satisfies $q_{0min}<q_{0}<1$. The expression for $q_{0min}$ is given by the (\ref{eq:q0minn}).

\begin{equation}
\label{eq:q0minn}
q_{0min}=\frac{\displaystyle\sum_{k=1}^{n}{\sum_{i=k}^{n}{k{n \choose i}{i \choose k} {q^{i}(1-q)^{n-i}}P_{0,i}^{k}\left(1-P_{d,i,0}\right)^{k}\left[1-P_{0,i}(1-P_{d,i,0})\right]^{i-k}}}}{\displaystyle\sum_{k=1}^{n}{\sum_{i=k}^{n}{k{n \choose i}{i \choose k} {q^{i}(1-q)^{n-i}}P_{0,i}^{k}\left(1-P_{d,i,0}\right)^{k}\left[1-P_{0,i}(1-P_{d,i,0})\right]^{i-k}}}+\sum_{k=0}^{n}{{n \choose k} {q^{k}(1-q)^{n-k}}P_{0d,k}}}
\end{equation}

Following the same methodology as in the two-user case, we obtain that the average queue size is given by:
\begin{equation}
\overline{Q}=\frac{\displaystyle(\lambda_{1}-\mu)\sum_{i=1}^{n}{i(i+3)p_{i}^{0}+\lambda_{0}\left(2\mu-\sum_{i=1}^{n}{i(i+3)p_{i}^{1}} \right)}}{2(\mu-\lambda_{1}+\lambda_{0})(\lambda_{1}-\mu)}
\end{equation}

The throughput per user for the network without the relay is given by
\begin{equation*}
\mu=\sum_{k=0}^{n-1}{{n-1 \choose k}q^{k+1}(1-q)^{n-1-k}P_{d,k+1}}
\end{equation*}
The throughput per user for the network with the relay is given by (~\ref{eq:mun}). The aggregate throughput is $\mu_{total}=n\mu$.

\begin{equation}
\label{eq:mun}
\begin{aligned}
\mu=q_{0}P\left(Q>0\right)\sum_{k=0}^{n-1}{{n-1 \choose k}q^{k+1}(1-q)^{n-1-k}P_{d,k+1,1}}+ \\
+\left[1-q_{0}P\left(Q>0\right)\right]\sum_{k=0}^{n-1}{{n-1 \choose k}q^{k+1}(1-q)^{n-1-k}\left[P_{d,k+1,0}+\left(1-P_{d,k+1,0}\right)P_{0,k+1}\right]}
\end{aligned}
\end{equation}

The throughput per user as a function of $q$ is given by (\ref{eq:muvsq}). In order to maximize $\mu(q)$, we need to find $q^*$ such that $\mu(q^*)\geq \mu(q)$ $\forall$ $0<q<1$. The analysis for finding the optimum is straight forward, has some complex calculations and will not add new insights to the results. We will present a numerical evaluation of this problem in the next section.

\begin{equation}
\label{eq:muvsq}
\begin{aligned}
\mu(q)=(1-q)^{n}\frac{\displaystyle\sum_{k=1}^{n}{\sum_{i=k}^{n}{A_{i,k}\left(\frac{q}{1-q}\right)^{i}}}}{\displaystyle\sum_{k=0}^{n}{{B_{k}\left(\frac{q}{1-q}\right)^{k}}+\sum_{k=1}^{n}\sum_{i=k}^{n}{A_{i,k}\left(\frac{q}{1-q}\right)^{i}}}}{\sum_{k=0}^{n-1}{{n-1 \choose k} \left(\frac{q}{1-q}\right)^{k+1}P_{d,k+1,1}}}+ \\
+(1-q)^{n}\left[1-\frac{\displaystyle\sum_{k=1}^{n}{\sum_{i=k}^{n}{A_{i,k}\left(\frac{q}{1-q}\right)^{i}}}}{\displaystyle\sum_{k=0}^{n}{{B_{k}\left(\frac{q}{1-q}\right)^{k}}+\sum_{k=1}^{n}\sum_{i=k}^{n}{A_{i,k}\left(\frac{q}{1-q}\right)^{i}}}} \right]{\sum_{k=0}^{n-1}{{n-1 \choose k} \left(\frac{q}{1-q}\right)^{k+1}\left[P_{d,k+1,0}+(1-P_{d,k+1,0})P_{0,k+1}\right]}}
\end{aligned}
\end{equation}

\begin{equation*}
\text{where }A_{i,k}=k{n \choose i}{i \choose k}P_{0,i}^{k}\left(1-P_{d,i,0}\right)^{k}\left[1-P_{0,i}(1-P_{d,i,0})\right]^{i-k}\\
\text{ and }B_{k}={n \choose k}P_{0d,k}
\end{equation*}

\section{Numerical Results}
\label{sec:results}
In this section we present numerical results for the analysis presented above. The results presented below have been verified by simulations which confirmed the accuracy of the analysis in the previous section. To simplify the presentation we consider the case where all the users have the same link characteristics and transmission probabilities. The parameters used in the numerical results are as follows. The distances in meters are given by $r(i,d)=r_{d}=130$, $r(i,0)=r_{0}=60$ $\forall i \geq 1$ and $r(0,d)=r_{0d}=80$. The path loss is $\alpha=4$ and the receiver noise power $\eta=10^{-11}$. The transmit power for the relay is $P_{tx}(0)=10$ mW and for the i-th user $P_{tx}(i)=1$ mW.

\subsection{Properties of the queue of the relay for the case of $n=2$ users}
\begin{figure}[ht]
\centering
\subfigure[Average Queue Size]{
\includegraphics[scale=0.4]{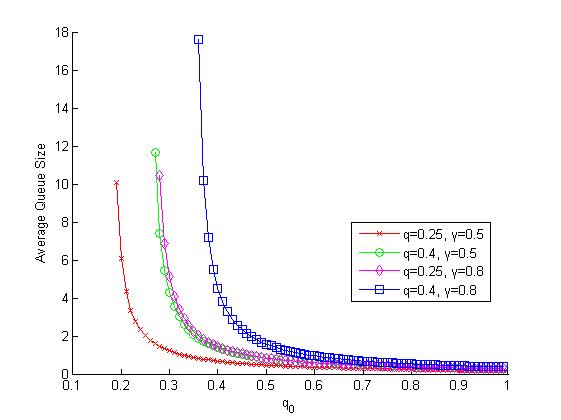}
\label{fig:avq}
}
\subfigure[Probability of the queue to be empty]{
\includegraphics[scale=0.4]{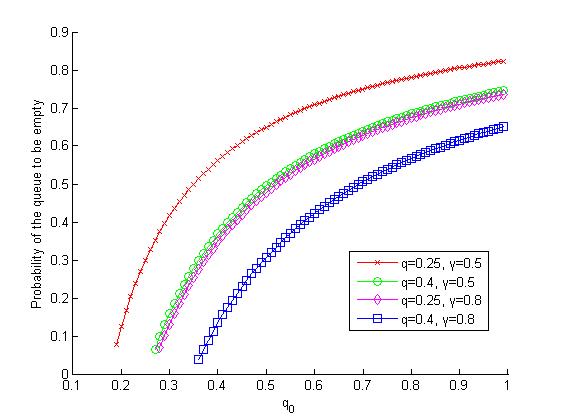}
\label{fig:prempty}
}
\label{fig:queuechars}
\caption{Properties of the relay's queue for the case of two-users}
\end{figure}

Fig.~\ref{fig:avq} and~\ref{fig:prempty} present the average queue size and the probability of the queue to be empty as the $q_{0}$ varies for various values of $q$ and $\gamma$. As the relay transmission probability $q_{0}$ increases then the queue is more likely to be empty. Equally expected is the decrease of the average queue size as $q_{0}$ increases.

\subsection{The impact of the number of users}
\begin{figure}[ht]
\centering
\subfigure[$\gamma=0.5$]{
\includegraphics[scale=0.4]{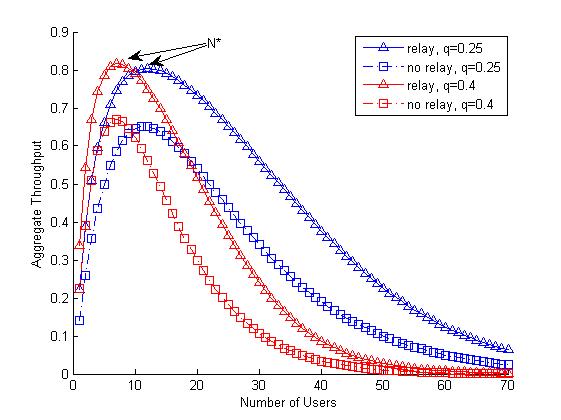}
\label{fig:ag_g1}
}
\subfigure[$\gamma=0.8$]{
\includegraphics[scale=0.4]{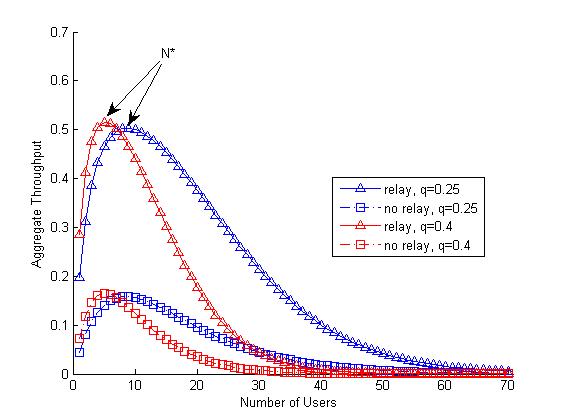}
\label{fig:ag_g2}
}
\label{fig:aggrthrlowg}
\caption{Aggregate throughput vs number of users for $\gamma < 1$}
\end{figure}

\begin{figure}[ht]
\centering
\subfigure[$\gamma=1.2$]{
\includegraphics[scale=0.4]{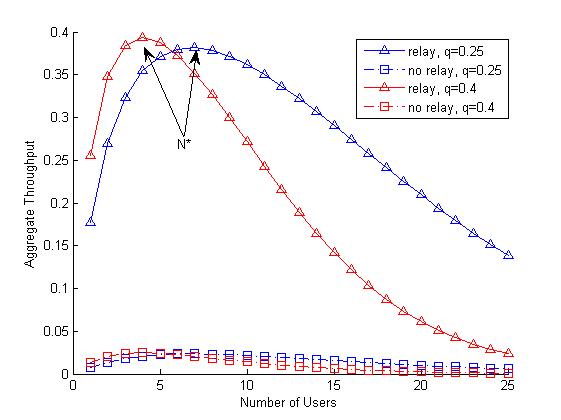}
\label{fig:ag_g3}
}
\subfigure[$\gamma=2.5$]{
\includegraphics[scale=0.4]{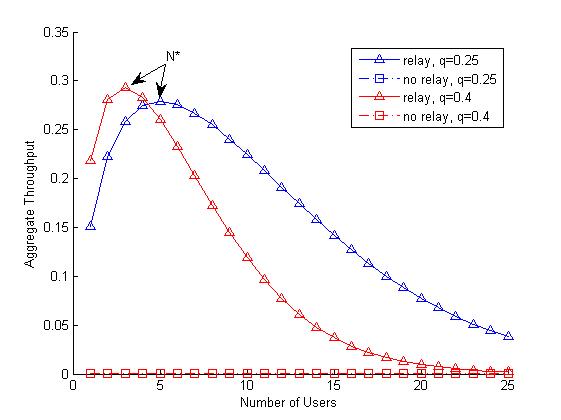}
\label{fig:ag_g4}
}
\label{fig:aggrthrhighg}
\caption{Aggregate throughput vs number of users for $\gamma > 1$}
\end{figure}

Fig.~\ref{fig:ag_g1} -~\ref{fig:ag_g2} and Fig.~\ref{fig:ag_g3} -~\ref{fig:ag_g4} show the aggregate throughput versus the number of users for $\gamma < 1$ and $\gamma > 1$ respectively. Notice that with small values of $\gamma$ is more likely to have more successful simultaneous transmissions comparing to larger $\gamma$. For $\gamma < 1$ it is possible for two or more users to transmit successfully at the same time, comparing to $\gamma > 1$ which that probability is almost zero.

The figures show that the relay offers a significant advantage compared to the network without the relay. When the threshold $\gamma$ increases the gain in term of percentage is greater. Another interesting observation is that given the link characteristics and the transmission probabilities, there is an optimum number of users $N^{*}$ that maximizes the aggregate throughput.  This number could be used as a criterion for finding the optimum size of a subset of users that a relay can serve.

\begin{figure}[ht]
\centering
\subfigure[$\gamma <1$]{
\includegraphics[scale=0.4]{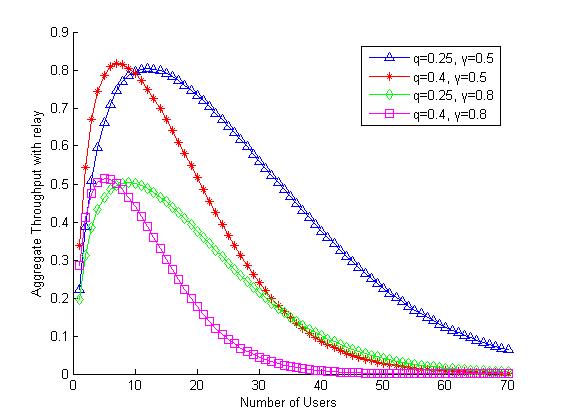}
\label{fig:aglowg}
}
\subfigure[$\gamma >1$]{
\includegraphics[scale=0.4]{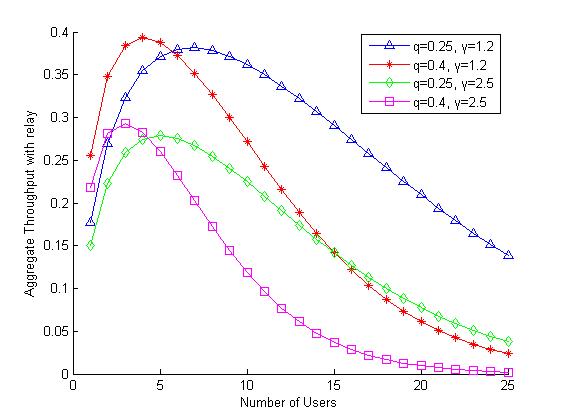}
\label{fig:aghighg}
}
\label{fig:aggthrrelay}
\caption{Aggregate throughput with relay vs number of users}
\end{figure}

Fig.~\ref{fig:aglowg} and~\ref{fig:aghighg} show the aggregate throughput versus the number of the users for several values of $q$ and $\gamma$. As $\gamma$ increases the number of users that achieves the maximum aggregate throughput is decreasing. The same conclusion comes for the values of $q$, as the $q$ increases.

\begin{figure}[ht]
\centering
\subfigure[$\gamma <1$]{
\includegraphics[scale=0.4]{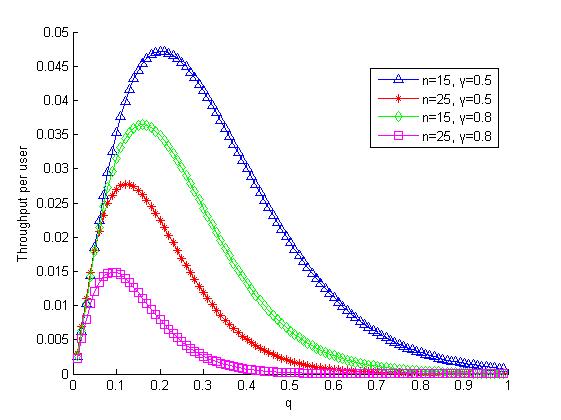}
\label{fig:thrvsqlow}
}
\subfigure[$\gamma >1$]{
\includegraphics[scale=0.4]{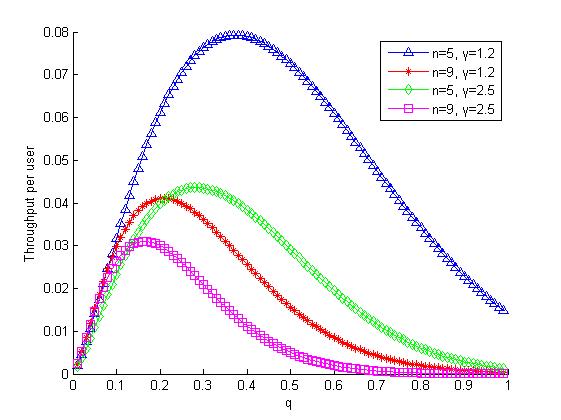}
\label{fig:thrvsqhigh}
}
\label{fig:thrvsq}
\caption{Throughput per user vs q}
\end{figure}

Fig.~\ref{fig:thrvsqlow} and~\ref{fig:thrvsqhigh} show the throughput per user versus the user's transmission probability $q$ for several values of $n$ and $\gamma$. An intuitive result for $q^{*}$ (the value of $q$ that maximizes the throughput per user), is that as $n$ increases then the $q^{*}$ decreases.

\begin{figure}[ht]
\centering
\subfigure[$\gamma <1$]{
\includegraphics[scale=0.4]{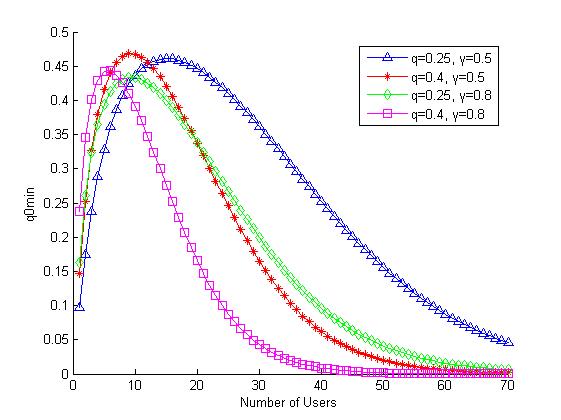}
\label{fig:q0minvsnlow}
}
\subfigure[$\gamma >1$]{
\includegraphics[scale=0.4]{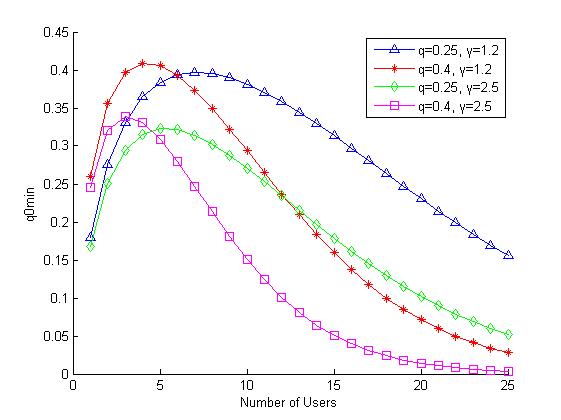}
\label{fig:q0minvsnhigh}
}
\label{fig:q0minvsn}
\caption{$q_{0min}$ vs Number of users}
\end{figure}

Fig.~\ref{fig:q0minvsnlow} and~\ref{fig:q0minvsnhigh} present the $q_{0min}$ threshold versus the number of users for $\gamma < 1$ and $\gamma > 1$ respectively.

The advantage that the relay offers is more obvious when the number of users is large. This is expected and feasible because of the MPR capabilities and the capture effect of the channel comparing to the collision channel in our previous work.

\section{Conclusions}
\label{sec:conclusions}
In this paper, we examined the operation of a node relaying packets from a number of users to a common destination node.  We assumed MPR capability for the relay and for the destination node. We studied a multiple capture model, where a user's transmission is successful if the received $SINR$ is above a threshold $\gamma$. We obtained analytical expressions for the relay's queue characteristics such as the stability condition, the values of the arrival and service rates, the average queue size. We showed that the arrival rate at the queue is independent of the relay probability of transmission, when the queue is stable. We studied the throughput per user and the aggregate throughput, and found that, under stability conditions, the throughput per user does not depend on the relay probability of transmission. The analytical results have been verified with simulations.
In Section~\ref{sec:results} we have given the conditions under which the utilization of the relay offers significant advantages. An interesting result is that, given the link characteristics and the transmission probabilities, there is an optimum number of users that maximizes the aggregate throughput. These results could be useful in a network with many users and multiple relays for determining the way to allocate the users among the relays. With the MPR and the capture effect the advantages from deploying a relay node are more pronounced.

An extension of the present work is the case of the relay node which is capable of transmitting and receiving packets at the same time (full duplex), the case of multiple relays (with possible cooperation among them) it is interesting too. Another possible extension is the case of dynamic adjustment of the transmission probabilities depending on the network conditions. Future extensions of this work will include users with non-saturated queues i.e. sources with external random arrivals, a relay node with its own packets and different priorities for the users. Another interesting extension of this work concerns the energy consumption in the total network, and in particular at the relay node.

\section*{Acknowledgements}
This research has been co-financed by the European Union (European Social Fund – ESF) and Greek national funds through the Operational Program ``Education and Lifelong Learning" of the National Strategic Reference Framework (NSRF) - Research Funding Program: Heracleitus II. Investing in knowledge society through the European Social Fund.

This work was also supported in part by MURI grant W911NF-08-1-0238, NSF grant CCF-0728966, and ONR grant N000141110127.

{
\section*{References}
\bibliographystyle{elsarticle-num}
\bibliography{bibliography}}

\begin{thebibliography}{10}
\expandafter\ifx\csname url\endcsname\relax
  \def\url#1{\texttt{#1}}\fi
\expandafter\ifx\csname urlprefix\endcsname\relax\def\urlprefix{URL }\fi
\expandafter\ifx\csname href\endcsname\relax
  \def\href#1#2{#2} \def\path#1{#1}\fi

\bibitem{b:Pappas-WiOpt}
N.~Pappas, A.~Ephremides, A.~Traganitis, Stability and performance issues of a
  relay assisted multiple access scheme with mpr capabilities, in: Modeling and
  Optimization in Mobile, Ad Hoc, and Wireless Networks, 2011. WiOPT 2011. 9th
  International Symposium on, 2011.

\bibitem{b:Pappas}
N.~Pappas, A.~Traganitis, A.~Ephremides, Stability and performance issues of a
  relay assisted multiple access scheme, in: Global Telecommunications
  Conference, 2010. GLOBECOM 2010. IEEE, 2010.

\bibitem{b:Muelen}
E.~C. V.~D. Meulen, Three-terminal communication channels, Advances in Applied
  Probability 3~(1) (1971) pp. 120--154.

\bibitem{b:CoverGamal}
T.~Cover, A.~Gamal, Capacity theorems for the relay channel, Information
  Theory, IEEE Transactions on 25~(5) (1979) 572 -- 584.
\newblock \href {http://dx.doi.org/10.1109/TIT.1979.1056084}
  {\path{doi:10.1109/TIT.1979.1056084}}.

\bibitem{b:Sadek}
A.~Sadek, K.~Liu, A.~Ephremides, Cognitive multiple access via cooperation:
  Protocol design and performance analysis, Information Theory, IEEE
  Transactions on 53~(10) (2007) 3677 --3696.
\newblock \href {http://dx.doi.org/10.1109/TIT.2007.904784}
  {\path{doi:10.1109/TIT.2007.904784}}.

\bibitem{b:Simeone}
O.~Simeone, Y.~Bar-Ness, U.~Spagnolini, Stable throughput of cognitive radios
  with and without relaying capability, Communications, IEEE Transactions on
  55~(12) (2007) 2351 --2360.
\newblock \href {http://dx.doi.org/10.1109/TCOMM.2007.910699}
  {\path{doi:10.1109/TCOMM.2007.910699}}.

\bibitem{b:Rong1}
B.~Rong, A.~Ephremides, Protocol-level cooperation in wireless networks: Stable
  throughput and delay analysis, in: Modeling and Optimization in Mobile, Ad
  Hoc, and Wireless Networks, 2009. WiOPT 2009. 7th International Symposium on,
  2009, pp. 1 --10.
\newblock \href {http://dx.doi.org/10.1109/WIOPT.2009.5291617}
  {\path{doi:10.1109/WIOPT.2009.5291617}}.

\bibitem{b:Rong2}
B.~Rong, A.~Ephremides, Cooperation above the physical layer: The case of a
  simple network, in: Information Theory, 2009. ISIT 2009. IEEE International
  Symposium on, 2009.

\bibitem{b:Pappas-ITW11-Relay}
N.~Pappas, A.~Ephremides, A.~Traganitis, Relay-assisted multiple access with
  multi-packet reception capability and simultaneous transmission and
  reception, in: 2011 IEEE Information Theory Workshop (ITW), 2011, pp.
  578--582.
\newblock \href {http://dx.doi.org/10.1109/ITW.2011.6089522}
  {\path{doi:10.1109/ITW.2011.6089522}}.

\bibitem{b:Pappas-ISIT12}
N.~Pappas, J.~Jeon, A.~Ephremides, A.~Traganitis, Wireless network-level
  partial relay cooperation, in: 2012 IEEE International Symposium on
  Information Theory Proceedings (ISIT), 2012, pp. 1122--1126.
\newblock \href {http://dx.doi.org/10.1109/ISIT.2012.6283028}
  {\path{doi:10.1109/ISIT.2012.6283028}}.

\bibitem{b:Pappas-arXiv-full-duplex}
N.~Pappas, M.~Kountouris, A.~Ephremides, A.~Traganitis, Relay-assisted multiple
  access with full-duplex multi-packet reception, in: arXiv:1310.2773 [cs.IT],
  2013.

\bibitem{b:Bertsekas}
D.~Bertsekas, R.~Gallager, Data networks (2nd ed.), Prentice-Hall, Inc., Upper
  Saddle River, NJ, USA, 1992.

\bibitem{b:AlohaVerdu}
S.~Ghez, S.~Verdu, S.~Schwartz, Stability properties of slotted aloha with
  multipacket reception capability, Automatic Control, IEEE Transactions on
  33~(7) (1988) 640 --649.
\newblock \href {http://dx.doi.org/10.1109/9.1272} {\path{doi:10.1109/9.1272}}.

\bibitem{b:Angel}
G.~del Angel, T.~L. Fine,
  \href{http://dx.doi.org/10.1109/TNET.2004.838605}{Optimal power and
  retransmission control policies for random access systems}, IEEE/ACM Trans.
  Netw. 12 (2004) 1156--1166.
\newblock \href {http://dx.doi.org/http://dx.doi.org/10.1109/TNET.2004.838605}
  {\path{doi:http://dx.doi.org/10.1109/TNET.2004.838605}}.
\newline\urlprefix\url{http://dx.doi.org/10.1109/TNET.2004.838605}

\bibitem{b:Naware}
V.~Naware, G.~Mergen, L.~Tong, Stability and delay of finite-user slotted aloha
  with multipacket reception, Information Theory, IEEE Transactions on 51~(7)
  (2005) 2636 -- 2656.
\newblock \href {http://dx.doi.org/10.1109/TIT.2005.850060}
  {\path{doi:10.1109/TIT.2005.850060}}.

\bibitem{b:Tse}
D.~Tse, P.~Viswanath, Fundamentals of wireless communication, Cambridge
  University Press, New York, NY, USA, 2005.

\bibitem{b:Nguyen}
G.~Nguyen, S.~Kompella, J.~Wieselthier, A.~Ephremides, Optimization of
  transmission schedules in capture-based wireless networks, in: Military
  Communications Conference, 2008. MILCOM 2008. IEEE, 2008, pp. 1 --7.
\newblock \href {http://dx.doi.org/10.1109/MILCOM.2008.4753605}
  {\path{doi:10.1109/MILCOM.2008.4753605}}.

\bibitem{b:NOW}
L.~Georgiadis, M.~J. Neely, L.~Tassiulas, {Resource Allocation and Cross-Layer
  Control in Wireless Networks}, Foundations and Trends in Networking, NOW,
  2006.

\bibitem{b:Gebali}
F.~Gebali, Analysis of Computer and Communication Networks, Springer, 2010.

\end{thebibliography}
\end{document}